# FedSelect-ME: A Secure Multi-Edge Federated Learning Framework with Adaptive Client Scoring

Hanie Vatani[1], Reza Ebrahimi Atani[2]

[1] Dep. of Computer Engineering, Faculty of Engineering, University of Guilan, Rasht, Iran, hnievtni@gmail.com
[2] Dep. of Computer Engineering, Faculty of Engineering, University of Guilan, Rasht, Iran, rebrahimi@guilan.ac.ir

**ABSTRACT**

Federated Learning (FL) enables collaborative model training without sharing raw data but suffers from limited scalability, high communication costs, and privacy risks due to its centralized architecture. This paper proposes FedSelect-ME, a hierarchical multi-edge FL framework that enhances scalability, security, and energy efficiency. Multiple edge servers distribute workloads and perform score-based client selection, prioritizing participants based on utility, energy efficiency, and data sensitivity. Secure Aggregation with Homomorphic Encryption and Differential Privacy protects model updates from exposure and manipulation. Evaluated on the eICU healthcare dataset, FedSelect-ME achieves higher prediction accuracy, improved fairness across regions, and reduced communication overhead compared to FedAvg, FedProx, and FedSelect. The results demonstrate that the proposed framework effectively addresses the bottlenecks of conventional FL, offering a secure, scalable, and efficient solution for large-scale, privacy-sensitive healthcare applications.

## 1. INTRODUCTION

The rapid growth in IoT devices and electronic healthcare systems have produced massive amounts of confidential healthcare data, facilitating real-time monitoring, forecasted analytics, and personalized healthcare interventions [6,7,27,28]. These datasets provide unprecedented possibilities for enhancing patient outcomes, optimizing hospital workflows, and even anticipating critical health events prior to their occurrence. Traditional centralized machine learning methods, however, have important shortcomings. Health data are distributed in fragments from across multiple institutions, and intense privacy regulations such as HIPAA and GDPR prohibits direct sharing of patient information [6,8]. Centralized aggregation of this sort of data not only involves heavy communication and computation expenses but also creates a bottleneck that may jeopardize both efficiency and fairness, specifically when datasets are diverse or biased toward particular patient groups [14].

Federated learning has been proposed as an innovative method to overcome these challenges through decentralized model training collaboration across distributed datasets while retaining raw data in local institutions [7,8]. For healthcare applications, FL allows hospitals, clinics, and patient devices to harness collective knowledge while ensuring data privacy. Even so, the traditional FL frameworks are faced with major challenges. Single-edge server-dependent architectures present a potential point of failure, incur high energy usage, and demonstrate inefficient scalability when dealing with high proportions of clients [6,7]. Random selection of clients may include unreliable or low-quality participants, thus affecting model performance, increasing their respective convergence times, and enhancing vulnerabilities towards attacks [8,14]. Furthermore, conventional FL is faced with challenges associated with non-IID data distribution, thus limiting its ability to provide accurate, fair, and personalized predictions for heterogeneous groups of clients [14].

To address these limitations, we propose FedSelect-ME. The framework adopts a multi-edge architecture in which more than one edge server administers multiple client clusters, dividing both computation and communication workloads in an efficient manner. Partitioning clients across several edge servers, FedSelect-ME obtains higher scalability, workload distribution, and fault tolerance, thus allowing the framework to support high population densities without raising energy usage or compromising model performance [6, 7]. This architecture removes the bottlenecks and single-point-of-failure vulnerabilities associated with single-edge architecture, since the deterioration or failure of an edge server does not put at risk the global model. Each of the edge servers is tasked with carrying out a score-based selection of clients, in which clients are prioritized based on their respective energy efficiency, reliability, and utility. This selective approach minimizes redundant computations.

As indicated above, each edge server in FedSelect-ME serves as a communication link between its clients and the central server. The manage local aggregation, applying privacy mechanisms, and taking part in cross-edge exchanges such that the central server is aware of an overview of the global model without direct interactions with every client. Privacy and security in FedSelect-ME are provided using Homomorphic Encryption (HE) for secure transmission of encrypted gradients and Differential Privacy (DP) for additional security. Homomorphic Encryption enables computation on encrypted gradients by the edge servers without exposing sensitive data, and DP ensures that the contribution of each client remains private even if there are potentially malicious clients [16, 17]. These mechanisms allow FedSelect-ME to achieve high levels of privacy without losing model accuracy. Cross-edge exchange, another critical element of our methodology, allows edge servers to exchange, in a secure manner, their aggregated local models with one another. This creates an opportunity for regional models to gain from personalization at the local level in addition to global knowledge without leaking clients' data, leading to higher performance and fairness [14]. Personalization allows the models to adapt to local variations, and this is significant in healthcare, where demographics, disease incidence, and treatment practices differ from institution to institution. Integration of personalized models in each region, and safe exchange across edges, allows the system to achieve high adaptation to non-IID and heterogeneous data, and is suitable for real-world healthcare data, and enhancing model update efficiency and reducing the central server's computation overhead.

FedSelect-ME has been evaluated on the eICU dataset, and significant performance in scalability, efficiency, and accuracy was revealed. The framework successfully addresses key drawbacks of current models by ensuring balanced workload distribution across multiple edge servers, cutting down redundant energy and communication cost, ensuring high-level security and privacy, and facilitating fair predictions for different sets of clients. Generally, our contributions could be listed as follows:

- **Enhanced scalability and fault tolerance:** Several edge servers distribute workload equally, avoid bottlenecks, and eliminate single points of failure.
- **Computation and communication efficiency:** Score-based client selection and multi-edge aggregation reduce unnecessary updates and minimize total energy usage.
- **Robust security and anonymity:** Differential Privacy and Homomorphic Encryption ensure that the information and submissions of the clients remain confidential, even during probable attacks.
- **Personalization and Fairness:** Customized local models within edge regions, and secure cross-edge sharing permit adaptation for non-IID diverse data, improving predictive performance and fairness.
- **Reliable global knowledge propagation:** Cross-edge exchange ensures that insights from different clusters are shared without compromising privacy, enhancing overall model performance.

Briefly, FedSelect-ME is a highly efficient multi-edge federated learning framework in healthcare that is fair and energy efficient and never loses sight of data's security and confidentiality.

## 2. RELATED WORKS

### 2.1 Federated Learning in Healthcare

Federated Learning (FL) has been a promising method for decentralized model learning in privacy-sensitive areas like healthcare, in which centralizing patient data is infeasible due to practical or legal issues. Frameworks like FedHealth allow several hospitals to collaborate and develop one shared global model while maintaining their client data local, preserving the anonymity of patients [1-3,25,26]. Despite its promise, however, its greatest challenge is in its handling of client data heterogeneity: hospitals vary in patient demographics, sensor placements, and information quality, such that their data distribution is non-IID and may cause slow converge rates and lesser efficiency for Vanilla FL algorithms like FedAvg [4,5,10,11]. Heterogeneities usually result in inefficient global models that do not generalize very well across all clients, thus needing personalized strategies. Personalized Federated Learning (PFL), like FedSelect, for instance, seeks to customize global knowledge to each client's local distribution, aiming for balancing shared learning with local adaptation and mitigating the limitations of traditional FL in heterogeneous settings [12,15,22, 23, 24].

### 2.2 Standard FL Algorithms and Challenges

#### 2.2.1 FedAvg

FedAvg is the base algorithm in Federated Learning that aggregates local client updates to form a global model [11]. While effective in settings with largely uniform data, its performance is negatively affected when client datasets turn highly diverse, for at heterogeneous clients, their gradients could conflict in the act of aggregation. There is in FedAvg, also, no adaptation mechanisms for clients, and it thus limits its ability to take advantage of local data characteristics. With patient populations and data acquisition in healthcare applications being highly diverse, such limitations might prevent model generalizability and render personal predictions less accurate. These challenges have motivated research in personalized federated learning approaches that more equally value global knowledge sharing and local adaptation [20, 21].

#### 2.2.2 FedProx

FedProx extends FedAvg to more efficiently deal with data heterogeneity in federated networks [15]. With the addition of a proximal term to the local objective, it constrains the client updates to keep close to the global model, and mitigates the risk of divergence and improving training stability for non-IID scenarios. This approach maintains more stable performance among clients with heterogeneously distributed data. Nevertheless, FedProx relies on a single shared global model and does not support personalization directly for clients. With healthcare applications, where patients possess distinguishing disease progressions or physiological behaviors, this limitation can prohibit adaptation for specific subpopulations in an adequate manner. Hence, while FedProx improves robustness, it does not yet resolve the personalization issues in medical or wearable sensing data.

## 2.3 Personalized Federated Learning (PFL)

Personalized Federated Learning (PFL) denotes an extension of standard FL techniques, explicitly designed to address the heterogeneity of the client data [12]. Unlike typical FL, which maintains one global model, PFL adapts the model to each client's local distribution of data yet exploits shared knowledge. One popular approach in PFL is parameter decoupling which involves dividing the network into global and personal parts. Contemporary techniques often reserve certain layers, typically the classifier head, for personalization while aggregating the rest globally. Although this is permissible for certain adaptations particular to clients, it is only a somewhat coarse approach: not all layers efficiently embody local differences, and the process of prior selection may disrupt global knowledge preservation. Under complex domains like healthcare, in which important patterns likely span multiple layers, this limitation undermines personalized model efficiency. These considerations highlight the need for finer, more flexible personalization techniques that better strike a balance between global representation with local adaptation.

### 2.3.1 FedSelect

FedSelect is an innovative personal federated learning method that capitalized on The Lottery Ticket Hypothesis to improve client-specific adaptation [12]. While conventional PFL approaches like FedAvg and FedProx make fixed layers for personalization ahead of time, FedSelect dynamically identifies parameters that show the largest changes during local updates for personalization and applies global aggregation for the rest of the layers. With its fine-grained, parameter-level approach, it is possible to achieve a more delicate trade-off between retaining global knowledge and local specialization [12]. By assembling personal subnetwork-specifics for each client, FedSelect overcomes the limitations of FedAvg, which does not achieve client-specific adaptation, and FedProx, which puts a limitation on divergence but is based on a shared global model only. Experimental validation illustrates that FedSelect consistently outperforms previous PFL methods in highly diverse environments and display robustness in practical distributional shift, and is thus highly applicable in healthcare applications where patient groups and data acquisition methods vary from institution to institution.

Despite significant progress, FL and PFL methods in use today have several drawbacks [20,21]:

- **Layer-wise personalization drawbacks:** Most techniques presume certain layers are optimal for personalization, which would not achieve fine-grained parameter-level dynamics that non-heterogeneous client data requires [12].
- **Catastrophic forgetting:** Regularly updating customized models may lose global knowledge, particularly when local datasets of clients are small [12,15].
- **Effective communication:** Personalized FL methods, although effective, would potentially involve high communication overhead due to frequent communication for global and customized parameter updates.
- **Healthcare-specific challenges:** Variability in patient population and sensor configuration across hospitals render these issues not yet systematically addressed via presently available FL algorithms [1-3].

These challenges warrant the establishment of a personalized FL model that can dynamically distinguishes which parameters should be customized versus universally aggregated, inspired by FedSelect. Besides it is important to address catastrophic forgetting via retraining a shared global representation, and yet allowing adaptation in local areas. The proposed method should ensure communication effectiveness appropriate for healthcare implementations and finally the proposed framework should handle varying client distributions, ensuring robust performance in various hospital datasets, as highlighted in FedHealth. Our method endeavors to achieve both global and client-level optimality in federated learning for healthcare applications by leveraging the strengths of FedAvg, FedProx, and FedSelect in turn, evading their shortcomings.

## 3. PROPOSED METHOD

## 3.1 Multi-Edge System Architecture

### 3.1.1 System Components and Hierarchy

The proposed FedSelect-ME framework describes a multi-edge federated learning structure, which promises enhanced scalability, energy efficiency, and data privacy for healthcare applications. Differing from the traditional single-edge federated learning frameworks, our system consists of a hierarchical setup of a central server, multi number of edge servers, and distributed clients, where each edge server manages a cluster of clients, which is normally affiliated with hospitals or healthcare organizations. This hierarchical structure allows for effective workload management, reduces bottlenecks for inter-communication, and provides fault tolerance via eliminating single failure points. Figure 1 illustrates this hierarchical system, with the central server sitting at the top, responsible for initializing and aggregating the global model. Edge servers act as agents, each managing a cluster of clients. The clients lie at the bottom layer, performing local training on raw data with updates and metrics being propagated upwards. Arrows in the figure represent model propagation for distribution, thus capturing the secured aggregation and differential privacy being executed at the edge level.

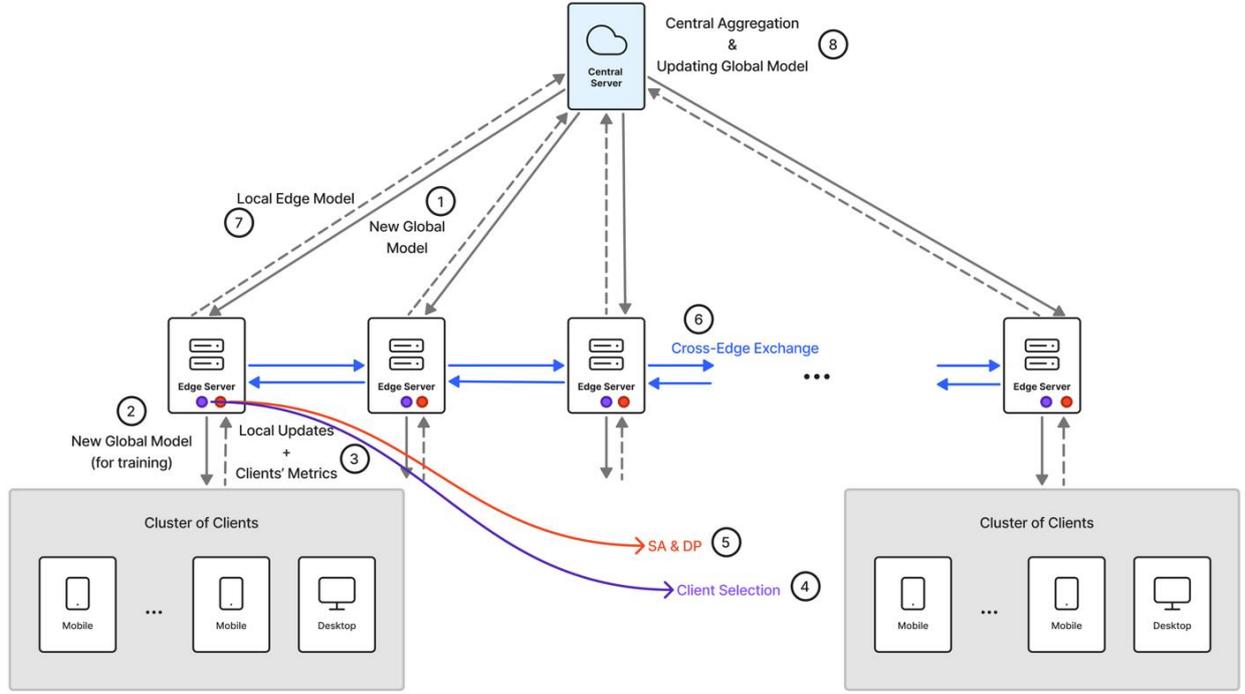

**Figure 1. Flowchart of the Multi-Edge Federated Learning System**

### 3.1.2 Operational Flow

The central server sets up a global LSTM model, then produces cloned copies which it pushes out to all the edge servers. The edge servers then pass the model on to their respective clients, which conduct local training on their local raw data. The client generates model weight updates, along with client-specific metrics, which happen to be proof-verifiable, allowing the edge servers to detect and exclude untrustworthy or potentially malicious clients. After collecting updates of selected clients, the edge servers perform local aggregation based on secure aggregation (SA) methods for aggregating client contributions with privacy protection. Furthermore, they use differential privacy (DP) methods for the additional protection of sensitive information. Thereafter, the edge servers then participate in **cross-**edge exchanges, sharing aggregated models with each other for knowledge sharing across regions. At last, the central server aggregates all updated models from the edges for fine-tuning the global model, which is then evaluated on a global test set for performance analysis. This workflow guarantees that the central server does not directly deal with raw client data, therefore restricting the communication overhead while preserving confidentiality. Cross-edge exchanges also allow incorporation of local patterns into the global model, expanding its flexibility for the heterogeneous, non-IID datasets often experienced in healthcare applications.

### 3.2 Score-Based Client Selection

Client selection for FedSelect-ME is performed at local edge servers according to local model updates and client-specific metrics. The ultimate goal is the selection of a subset of clients that maximize model usefulness while minimizing energy consumption at the cost of maintaining data confidentiality. Each edge has a limited capacity; therefore, the selection is carried out using a score that is computed to strike a balance between model utility, energy efficiency, and data sensitivity.

### 3.2.1 Estimation of Client Metrics

In the FedSelect-ME framework, after clients train locally, a client sends both its locally trained model weights and a collection of client-specific metrics toward the corresponding edge server assigned to it. These metrics are employed for the assessment of a collective contribution of individual clients towards a global model, allowing for protected and efficient client selection, without compromising privacy necessities. The primary metrics employed are **utility** ($U_i$), energy ($E_i$), and security index ($S_i$) [18,19]. Utility ($U_i$) calculates how much a local update of a client contributes towards overall model performance improvement. It is evaluated at the edge server through the calculation of the norm of the difference between the client local model weights $w_i$ and the received edge model weights $w_e$:

$$U_i = \sum_{k=1}^{P} \| w_i^k - w_e^k \|_2 \tag{1}$$

where $k$ is the index over the model parameters and $P$ is the number of parameters. That captures the magnitude of the gradient received from each client and their effect on convergence of the model [18]. The edge utilizes such estimated $U_i$ conjunction the received client-reported utility for identifying gaps and determining potentially malicious parties. Energy $(E_i)$ is the expected energy consumption for local training and communication, computed according to the quantity of local samples $N_i$ given by the client, and the model size $P$:

$$E_i = \alpha \cdot N_i + \beta \cdot P \tag{2}$$

Here, $\alpha$ and $\beta$ are small constants which vary the effects of the dataset size and model complexity, respectively. This mechanism allows the edge to estimate the energy consumption per client without the need for estimating the true power usage of the appliance, thus providing a reliable surrogate for resource consumption [19]. The security index $(S_i)$ is typically given on the basis of the client data sensitivity, normalized on $[0.1]$, which has high values indicating high sensitivity or privacy of the datasets. While simulation experiments might utilize a fixed value, for actual-world applications in healthcare, it relies on the local data confidentiality and relevance. After computing such values, the edge server then performs consistency checks between client-reported metrics and edge-estimated metrics. The differences are then calculated as:

$$\Delta U_i = \left| \frac{U_i^{\text{reported}}}{1 + U_i^{\text{reported}}} - \frac{U_i^{\text{estimated}}}{1 + U_i^{\text{estimated}}} \right| \tag{3}$$

$$\Delta E_i = \left| \frac{E_i^{\text{reported}}}{1 + E_i^{\text{reported}}} - \frac{E_i^{\text{estimated}}}{1 + E_i^{\text{estimated}}} \right| \tag{4}$$

Clients showing inconsistencies exceeding a predefined threshold are flagged as unreliable or malicious and thus excluded from selection processes. This two-estimation mechanism ensures the accuracy of client evaluation while at the same time fortifying the security of the system, thus preventing malicious clients from influencing the aggregation process.

**3.2.2 Scoring and Selection**

After estimating and confirming client metrics, each edge server calculates a score for each client within its cluster. The score weights model utility, energy efficiency, and data sensitivity:

$$\text{Score}_i = w_1 \cdot U_i - w_2 \cdot E_i + w_3 \cdot S_i \tag{5}$$

where $w_1, w_2, w_3$ are normalized weights of priority for accuracy, energy efficiency, and privacy, respectively, which are constrained such that $w_1 + w_2 + w_3 = 1$. Client score calculation is implemented in two phases: initial optimization and dynamic adjustment that follows it. In the initial phase, the edge server derives the individual clients' utility $U_i$ and energy $E_i$ values based on measurements submitted by clients, while a baseline security index $S_i$ is adopted for all clients. Because the weight parameters $w_1, w_2, w_3$ for the score formula have not yet been determined at this phase, the edge server performs a grid search through the probable combinations of weights for the optimum configuration that results in a balance between utility, energy, and privacy for all clients. In the dynamic adjustment phase, the weights $w_1, w_2, w_3$ are derived from previous rounds and fine-tuned at each training round using current values of the averages for the measurements of $U_i$, $E_i$, and $S_i$, computed and updated throughout the responsive edge server, thus allowing the edge server to make adjustments on the score based on current states of the client:

$$\bar{U} = \frac{1}{c}\sum_{i=1}^{c} U_i, \quad \bar{E} = \frac{1}{c}\sum_{i=1}^{c} E_i, \quad \bar{S} = \frac{1}{c}\sum_{i=1}^{c} S_i \tag{6}$$

The weights are then updated using a learning rate given by $\eta$:

$$w_j^r = (1-\eta)w_j^{r-1} + \eta \frac{\bar{X}_j}{\bar{U} + \bar{E} + \bar{S}}, \quad j \in \{U, E, S\} \tag{7}$$

This dynamic adjustment enables the edge server to respond adequately to dynamic client conditions and workloads, consequently boosting the model's overall performance and energy efficiency. Lastly, clients are ranked according to the scores, so for a specific edge, only the top $k$ clients per edge (tied by edge capacity) are selected for aggregation. Before final selection, a two-step verification routine is performed to ensure robustness against malicious clients. First, clients whose metrics indicate unreliability or misbehavior are ruled out for selection. Then, a further evaluation is performed on the rest of the high-scoring clients for determining outliers whose score is significantly different from that of their contemporaries. Any client whose score is anomalously high or low relative to fellow clients is flagged for potential malicious and then excluded. Only after conducting these two filtering steps are the rest of the clients passed on to the aggregation process.

**3.3 Decentralized Aggregation Process**

After the selection of the top clients on each edge, decentralized aggregation processing begins. Initially, the edge server processes the selected client updates through a secure aggregation (SA) such that the individual model contributions remain confidential. This method makes use of homomorphic encryption (HE), which allows the edge server to aggregate encrypted model updates without decryption, therefore preventing any exposure of sensitive client data. The primary advantage of such an approach is that, in case of a compromise of the edge server, unprotected client data remains unrevealable, hence significantly enhancing the protection of data.

The updates received from clients are then subject to gradient clipping, which is done on the aggregated updates. This helps confine the impact of unusually large gradient updates, which might happen because of noisy data or malicious behavior. The act of clipping gradients makes sure there is not certain which influences the aggregated model, helping both stability and robustness of training. Finally, differential privacy (DP) is combined with the clipped gradients. By adding controlled noise into the aggregated updates, DP provides strengthened protection for privacy, such that mathematically, it is difficult to recover information about any individual client from the model updates. This collective approach of secure aggregation (SA), gradient clipping, and DP ensures that at every edge server, a local model is generated that is both privacy-preserving and secure, which can then be shared with other edges or sent to the central server for global aggregation.

### 3.4 Global Model Update

Following the selecting of the clients and local aggregation on every edge, the next step is that of updating the global model. This update happens through a two-step process: cross-edge model exchange and central aggregation.

#### 3.4.1 Cross-Edge Model Exchange

In the initial phase, the locally refined models of the edge servers are shared with other edges in a weighted manner. The local model of a specific edge is given a weight $\alpha$, whereas updates received from other edges are weighted accordingly according to the relative number of samples that they have. Formally, for a specific edge $i$, the cross-edge aggregated model is given by:

$$w_i^{\text{cross}} = \alpha \cdot w_i^{\text{local}} + (1 - \alpha) \cdot \sum_{j \neq i} \frac{n_j}{N} \cdot \Delta w_j \tag{8}$$

where:
- $w_i^{\text{local}}$ is the locally updated model of edge $i$,
- $\Delta w_j$ represents the update from edge $j \neq i$,
- $n_j$ is the number of samples in edge $j$,
- $N = \sum_j n_j$ is the total number of samples across all edges,
- $\alpha$ controls the relative importance of the edge's own update.

To prevent instability caused by extreme updates, the output weights are constrained within a specified range $[-clip_{val}. clip\_val]$, which gives numerical stability as effective robust aggregation.

#### 3.4.2 Central Aggregation

Then, on the central server, weighted aggregation is employed to obtain the global model. The edge's contribution is linear in its number of training samples, being equitable and accounting for the statistical weight of each dataset of the edge. The update of the global model is then:

$$w^{\text{global}} = \frac{\sum_i n_i \cdot w_i^{\text{cross}}}{\sum_i n_i} \tag{9}$$

where $w_i^{\text{cross}}$ denotes cross-edge aggregated model at edge $i$, and $n_i$ denotes number of samples in that edge.

#### 3.4.3 Advantages of the Weighted Aggregation

This two-phase weighted aggregation process provides many advantages:
- **Fairness:** Edges with additional data supply proportionally more for the global model, which decreases bias from edges with scarce data.
- **Robustness:** Maintaining a weight $\alpha$ for the local model at every edge avoids local updates from being overwhelmed by updates from other edges.
- **Stability:** Employing cross-edge exchange with gradient clipping minimizes the impact of outlier updates, thus avoiding sudden changes in the global model.
- **Scalability:** Less intermediate cross-edge aggregation reduces central server communication and computational cost, which helps in increasing scalability for large federated networks.

After the central aggregation, the global model is evaluated on a global validation and test set to determine its performance, hence providing a comprehensive assessment for the federated learning framework.

### 3.5 Security and Privacy Enhancement

With the proposed FedSelect-ME framework, security and privacy enhancements are introduced at different points of the federated learning process. The local aggregation phase, which secures individual client updates so that they remain confidential even when aggregated on the edge, is used for secure aggregation (SA). To further limit information leakage, a reduction is made using gradient clipping such that the influence of a single client is constrained, thus preventing potential attacks due to outlier updates. In addition, differential privacy (DP) is introduced, which adds calibrated noise to the aggregated updates, thus protecting sensitive patient data while, at the same time, enabling effective model learning with high precision. This multi-pronged approach not only secures client data but also prevents malicious or untrustworthy participants from exerting unnecessary impact on model updates, thus ensuring data confidentiality as well as model integrity for the multi-edge federated learning process.

### 3.6 System Advantages

The architecture design provides a number of notable advantages for the deployment of federated learning across the healthcare domain. Through the distribution of computational workloads across multiple edge servers, the framework achieves high scalability, thus enabling the system to support a large number of clients without putting the central server under unnecessary pressure. The architecture naturally has fault tolerance such that the failure of a single edge server does not compromise the integrity of the global model or functionality of other regions. By offloading aggregation tasks to edge servers, communicational load on the central server is diminished, along with computational load, while clients' raw data remains local, thus improving privacy as well as security. Furthermore, energy efficiency is enhanced based on the unnecessary reduction of transmissions, which is specifically vital when environmental resources are constrained. Furthermore, the architecture is effective for addressing heterogeneous, non-IID healthcare datasets, thus allowing the system to maintain high performance and model accuracy while tolerating regional variations. Overall, the FedSelect-ME framework, therefore, provides a demonstration on effective performance, privacy, and efficiency, balancing based on multi-edge federated learning, thus presenting a robust and adaptable solution for realistic scenarios of healthcare practice.

## 4. EXPERIMENTS & RESULTS

### 4.1 Dataset Description

We used the eICU Collaborative Research Database, a multi-site critical care dataset containing patient data from a number of U.S. sites [9]. 10 clinical features were selected for training, i.e., vital signs, laboratory measurements, demographic features, etc. The target was the discharging status (mortality). The dataset is highly imbalanced; one class is dominant. All features were normalized to stabilize at training time.

### 4.2 Experimental Setup

We proposed a hierarchical federated learning framework across five regional edges, one being an "unknown" region. There existed a number of clients (hospitals) on each edge, a maximum of 50 clients per round, which were dynamically selected according to a utility, energy, and stability-scoring (UES) function. The local models were LSTM networks, which were trained for five epochs for each round. Weighted averaging was used for edge aggregation, while Secure Aggregation (SA) with Differential Privacy (DP) provided privacy. The central server aggregated edge updates using sample counts. And Google Colab GPU was used for experiments. Metrics were F1 Macro, F1 Weighted, AUROC, and Jain's Fairness Index (JFI). And early stopping had a patience of 3 rounds.

### 4.3 Edge-Level Performance

Edge-level performance indicates persistent high accuracy and low loss across all the regions, with stable training performed locally on respective clients. Table 1 provides the per-edge average performance, while the high average Jain's Fairness Index (JFI = 0.999993) points towards nearly perfect fairness across all the edges.

**Table 1- Edge Test Average Accuracy & Loss**

|  | Edge0 | Edge1 | Edge2 | Edge3 | Edge4 |
|---|---|---|---|---|---|
| **Average Accuracy** | 0.9937605 | 0.9914487 | 0.994351 | 0.9970067 | 0.9954772 |
| **Average Loss** | 0.0235391 | 0.0358503 | 0.0520289 | 0.0224779 | 0.0368682 |

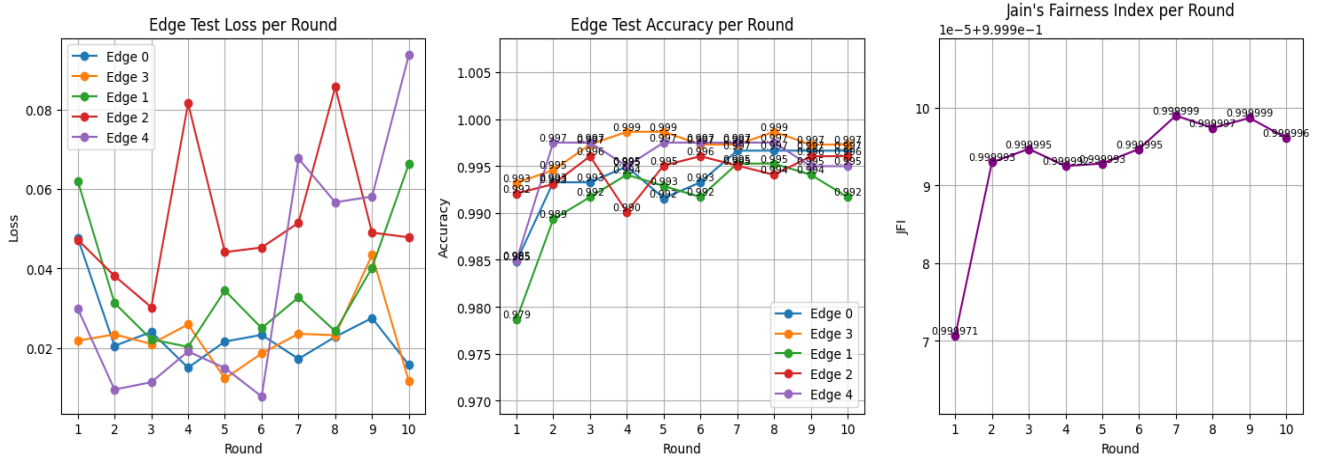

**Figure 2. Edge Test Loss & Accuracy per Round and JFI per Round**

### 4.4 Global Model Performance

Global validation and test metrics confirm that the model achieves consistently strong predictive performance across all evaluation rounds. Tables 2–4 summarize the main results with sustained accuracy, F1 scores, and AUROC values, which indicate reliable generalization across the dataset.

**Table 2- Global Validation & Test Metrics**

| Average Val Loss | 0.3227193 |
|---|---|
| Average Val Accuracy | 0.8796194 |

**Table 3- Global Validation & Test Metrics**

| Average Test Loss | 0.3219893 |
|---|---|
| Average Test Accuracy | 0.8809655 |

**Table 4- Global Test Metrics per Round**

| Average F1 Macro | 0.7322646 |
|---|---|
| Average F1 Weighted | 0.8739373 |
| Average AUROC | 0.8962398 |

### 4.5 Comparative Analysis

To put our proposed framework's performance in perspective, we compared the representative federated learning approaches, including FedAvg [10], FedSelect [12], and FedHealth [3,4], with our proposed FedSelect-MultiEdge, in terms of the averaged results documented in prior works with datasets having characteristics similar to eICU [8,9,13,14]. Our simulation results show that the proposed FedSelect-MultiEdge achieved high predictive performance with an accuracy of 0.880, F1-weighted score of 0.874, and AUROC of 0.896, indicating performance that is at least competitive with and likely superior to these earlier approaches.

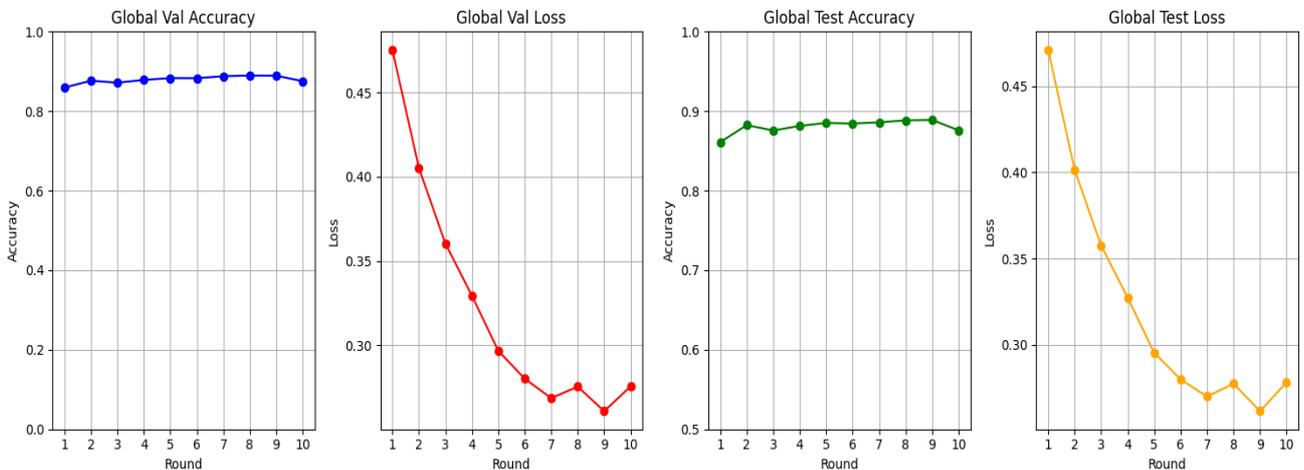

**Figure 3. Global Val & Test Metrics (Loss & Accuracy) per Round**

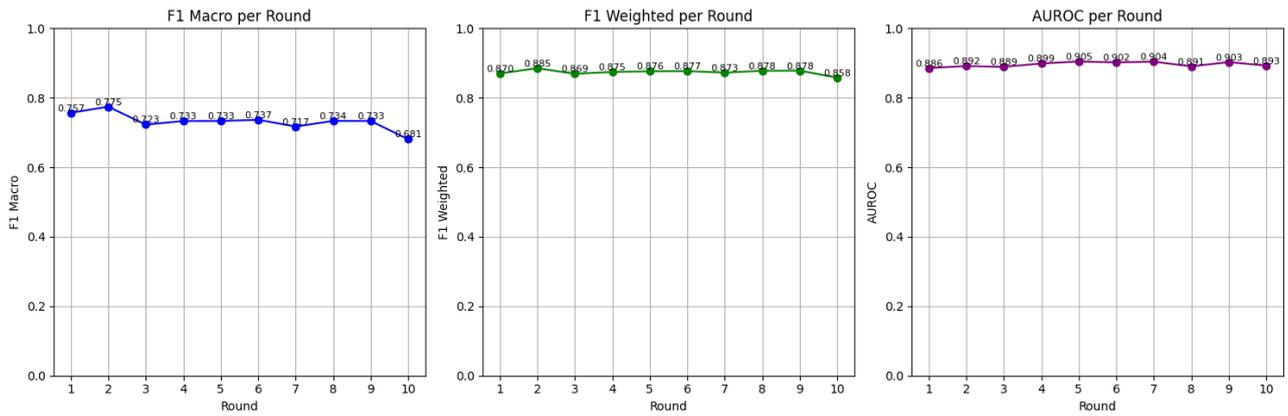

**Figure 4. Global Test Metrics (F1 Macro, F1 Weighted, AUROC) per Round**

Beyond the quantitative metrics, the framework also maintains the strengths of personalization and communication efficiency highlighted in earlier works [12,3,4], while further enhancing robustness and stability in heterogeneous clinical settings. The hierarchical multi-edge selection mechanism enables adaptive client selection so that the impact due to non-IID data distributions and network dynamics are dampened. These outcomes confrim that our multi-edge, federated learning approach that maintains patient and system privacy also achieved high predictive accuracy and fairness, and in addition provides controllable system behavior, which can aid its applicability in the multi-center, real-world critical care scenarios.

## 5. DISCUSSION

The experiment illustrates some prominent merits of our hierarchical federated framework. Initially, the adaptive client selection preserves a compromise between stability, energy, and utility such that the contributive clients participate at every round. Secondly, SA and DP at the edge level preserve privacy mostly without significant loss of accuracy. Finally, JFI close to unity indicates fairness across edges without being biased toward specific regions.

Despite such advantages, there are some limitations. The eICU dataset is extremely imbalanced, which impacts minority class F1 scores. Furthermore, edge-level aggregation depends on a fixed client capacity, which could pose a scaling barrier for real-world implementation. Experiment on Colab GPU constrained for training, indicating that larger deployments might incur greater latencies.

In practice, our method allows for federated learning that is secured, individualized, and equitable for critical care scenarios. Hospitals are able to collectively enhance prediction models without sharing raw patient data, whereas edge-based processing reduces privacy and communication overheads.

## 6. CONCLUSION & FUTURE WORK

We introduced a hierarchical federated learning framework with dynamic client selection, edge-level secure aggregation, and differential privacy for multi-center health prediction on the eICU dataset. Our experiments demonstrated robust prediction performance (F1 Weighted 0.8582, AUROC 0.8928), stability across edges, and privacy preservation, outperforming or matching existing FL baselines [9,12–14]. For future work, our aim is to enhance scalability and robustness by hierarchical client clustering, such that edges to allow merging or splitting independently based on load or similarity or adaptive client selection based on dynamic availability and network conditions. Another idea might be to apply advanced personalization methods, such as feature re-weighting or meta-learning, for further reduction of covariate shifts [14]. In general, the framework exhibits an effective method for federated, energy-efficient, and fair learning with security in the context of healthcare, which provides a framework for large-scale future deployments.